\documentclass[a4paper,12pt]{article}
\usepackage{amsmath}
\usepackage{amssymb}
\usepackage{tabularx}
\usepackage{enumerate}
\usepackage{graphicx}
\usepackage{subfigure}
\usepackage{color}
\usepackage[pagewise]{lineno}
\usepackage{longtable}
\usepackage{epstopdf}
\usepackage{float}
\usepackage{setspace}

\usepackage[justification=centering]{caption}

\usepackage{graphicx}
\usepackage{subfigure}
\newtheorem{thm}{Theorem}[section]
\newtheorem{lem}[thm]{Lemma}
\newtheorem{cor}[thm]{Corollary}

\newtheorem{rmk}{Remark}[section]

\newtheorem{pppp}{Proof}

\newcommand{\qed}{\hspace{1em}\mbox{\raisebox{0.65ex}{\fbox{}}}}

\numberwithin{equation}{section}

\newcommand{\be}{\begin{equation}}
\newcommand{\ee}{\end{equation}}
\newcommand\bes{\begin{eqnarray}}
\newcommand\ees{\end{eqnarray}}
\newcommand{\bess}{\begin{eqnarray*}}
\newcommand{\eess}{\end{eqnarray*}}

\newcommand{\R}{\mathbb{R}}
\newcommand{\bpf}{{\bf Proof:\ \ }}
\newcommand{\epf}{\mbox{}\hfill $\Box$}

\begin{document}
\thispagestyle{empty}

\title{Extinction, persistence and growing in a degenerate logistic model with impulses \thanks{The second author is supported by the National Natural Science Foundation of China (No. 12271470) and Carlos Alberto Santos acknowledges the support of CNPq/Brazil Proc. $N^{o}$ $311562/2020-5$ and FAPDF grant 00193.00001133/2021-80.}}
\date{\empty}

\author{Willian Cintra$^1$, Zhigui Lin$^{2,\dag}$, Carlos Alberto Santos$^{1}$ and Phyu Phyu Win$^{2}$\\
{\small $^1$ Department of Mathematics, University of Brasilia, BR-70910900 Brasilia, DF, Brazil}\\
{\small $^2$ School of Mathematical Science, Yangzhou University, Yangzhou 225002, China}
}

 \maketitle

\begin{quote}
\noindent {\bf Abstract.}
This paper deals with an impulsive degenerate logistic model, where pulses are introduced for modeling interventions or disturbances, and degenerate logistic term may describe refugees or protections zones for the species. Firstly, the principal eigenvalue depending on impulse rate, which is regarded as a threshold value, is introduced and characterized. Secondly, the asymptotic behavior of the population is fully investigated and sufficient conditions for the species to be extinct, persist or grow unlimitedly are given. Our results extend those of well-understood logistic and Malthusian models. Finally, numerical simulations emphanzise our theoretical results highlighting that medium impulse rate is more favorable for species to persist, small rate results in extinction and large rate leads the species to an unlimited growth.

 \medskip


\end{quote}

\section{Introduction}

This paper characterizes the dynamics of the impulsive and diffusive logistic-model possibly in the presence of a refugee or protection zones
\begin{eqnarray}
\left\{
\begin{aligned}
&u_t=d \Delta u+ a(t,x) u-b(t,x)u^p, && t\in((n\tau)^+,(n+1)\tau],\,\,x \in \Omega,  \\
&u(t,x)=0, && t>0,\,\,x \in \partial \Omega,\\
&u(0,x)=u_0(x), && x\in \overline \Omega,\\
&u((n\tau)^+,x)=cu(n\tau,x),&& x\in \overline \Omega,\ n=0,1,2,\cdots,
\end{aligned} \right.
\label{a01}
\end{eqnarray}
where $\Omega$ is a connected bounded domain of $\R^N$ ($N\geq 1$) with smooth boundary $\partial \Omega$, $p>1$ and $\tau>0$ are constants, $t\in((n\tau)^+,(n+1)\tau]$ expresses that the equation holds for $t\in((n\tau),(n+1)\tau]$, and the unknown $u$ takes its value $u((n\tau)^+,x)$ instead of $u((n\tau),x)$ at the initial time of the time interval $((n\tau),(n+1)\tau]$ for $n=0,1,2,\cdots$.

All meanings of variables and parameters in model \eqref{a01} are given below:

$\bullet$ $u(t,x)$ expresses the density of species at time $t \geq 0$ and in space $x \in \Omega$;

$\bullet$ $d>0$ is the diffusion rate;

$\bullet$ $a(t,x)(\in C^{\theta/2, \theta}([0, \tau]\times \overline \Omega)$ for some $0<\theta \leq 1$) is a $\tau-$ periodic function, and denotes the intrinsic growth rate of the population. It can be negative, which means that the resources on position $x$ at time $t$ are not conducive to survival;

$\bullet$ $b(t,x)(\in C^{\theta/2, \theta}([0, \tau]\times \overline \Omega))$ is a nonnegative and $\tau$-periodic function that can vanish on a smooth subdomain $\Omega_0$ of $\Omega$ (under additional assumptions on $b$, the subset $b^{-1}(0)$ may behave as periodic-parabolic protection or refugee zone, see below);

$\bullet$ $cu$ is the impulsive function, and an impulse occurs at every time $t=n\tau$ $(n=0,1,2,\cdots)$ within the successive stage of growth and disperse. Usually, the linear impulse function is used to represent birth pulse if $c>1$ and harvesting phenomenon if $0<c<1$.

On the condition $c=1$, which means there is no impulse, problem \eqref{a01} has attracted much attention recently, see \cite{AL2017,DLG2018,DH1999,H, LG2020} and references therein. Especially, a degenerate diffusive logistic equation was analyzed in \cite{CLT2023} from a numerical point of view and revealed that the number of nodal solutions and the number of components in the bifurcation diagrams strongly
depends on the number and position of the components where the weight function in front of the nonlinearity vanishes.

Recently, the single species model has also been extended to two species model, see for example \cite{LM2024}, that analyzed a spatially heterogeneous diffusive predator-prey model constructing an $S$-shaped component of coexistence states, which causes the existence of, at least, three coexistence states.

Corresponding elliptic problems to \eqref{a01}, without impulse and its variants, were well studied, see
\cite{DG2003, DG2006} and references therein, while corresponding periodic-parabolic problems has also attracted much attention, see for instance \cite{DP2012, LG2020} and references therein. Recently, it was proved in \cite{Sun2022} the existence of positive periodic solutions for a periodic-parabolic problem revealing that the linear part of reaction term and the nonlinear part make quite different effects on the dynamical behavior of positive periodic solutions.

Motivated by the above results, besides growth, death, and disperse \cite{OL2001}, we are more interested in the distribution and dynamics of species influenced by impulsive perturbation \cite{BZ2020, MLP2021,LZY2011, WZ2019} on the presence of a degenerate diffusive logistic term.

This paper is organized as follows. Section 2 contains some necessary theoretical results about the principal eigenvalue for a periodic eigenvalue problem with impulse. In Section 3, we establish a global existence result in the parameter $c>0$ for an impulsive periodic problem associated to problem \eqref{a01}. Sections 4 is devoted to studying dynamics of the unique solution of problem \eqref{a01}. Finally, numerical simulations  and a ecological explanations are given in Section 5.

\section{\bf An eigenvalue problem with impulse condition}

\mbox{}

Our main result about the distribution and the dynamic of the population (Theorem \ref{t41} below) strongly depends on the behavior of the principal eigenvalue  of the periodic $\mu$-eigenvalue problem
\begin{eqnarray}
\left\{
\begin{aligned}
&\phi_t-d \Delta \phi-a(t,x)\phi+ yb(t, x)\phi=\mu \phi, && t\in(0^+, \tau],\,\,x \in \Omega,  \\
&\phi(t,x)=0, && t\in [0, \tau],\,\,x \in \partial \Omega,\\
&\phi(0,x)=\phi(\tau, x), && x\in \overline \Omega,\\
&\phi(0^+, x)=z\phi(0, x),&& x\in \overline \Omega,
\end{aligned} \right.
\label{a013}
\end{eqnarray}
for any $z>0$ given.

The existence of the principal eigenvalue
$$\mu_1=\mu_1(d, a(t,x), yb(t,x), z),$$
for any $z>0$ and $y \geq 0$ given, and fixed $d, a(t,x)$ and $b(t,x)$,
can be guaranteed by using Poincar\'{e} map of a linear operator \cite{LZ, LZZ} or by using Krein-Rutman theorem \cite{H, KR} on a Banach space involving impulse \cite{XLS2023}.  For the convenience of the reader, we will give a sketch of the proof below.

To overcome the difficulties introduced by the impulse condition, we consider the following equivalent eigenvalue problem
\begin{eqnarray}
\left\{
\begin{array}{lll}
\xi_t-d \Delta \xi-a(t,x)\xi+yb(t,x)\xi=\mu\xi,& t\in(0,\tau],\,\,x \in \Omega,  \\[1mm]
\xi(t,x)=0, & t\in [0, \tau],\,\,x\in \partial\Omega,\\[1mm]
\xi(0,x)=z \xi(\tau,x), & x\in \overline \Omega,
\end{array} \right.
\label{3.02}
\end{eqnarray}
under the understanding that $\phi(t,x)=\xi(t,x)$ for $t\in (0,\tau],x\in \overline \Omega$, $\phi(0^+,x)=\xi(0,x)$ for $x\in \overline \Omega$, and $\phi(0,x)=\xi(\tau,x)$ for $x\in \overline \Omega$.

Now, let $W$ be a Banach space
\begin{eqnarray*}
\begin{array}{rll}
W=&{D}^{0,1}_0([0, \tau]\times \overline \Omega):=\{\xi(t,x)\in {C}^{0,1}([0, \tau]\times \overline \Omega): \ \xi(t,x)=0\ \forall (t,x)\in [0, \tau]\times \partial \Omega, \\
&\xi(0,x)=c\xi(\tau,x),\ \forall x\in \overline \Omega\}
\end{array}
\end{eqnarray*}
with its positive cone
$$
 {W^+}:=\textrm{closure}\{\xi\in W:\, \xi(t,x)\gg 0\ \forall (t,x)\in [0, \tau] \times \partial \Omega \},
$$
and its interior
$$
\textrm{Int} (W^+)=\{\xi\in W:\ \xi(t,y)\gg 0\ \forall (t,x)\in [0, \tau] \times \partial \Omega \}
$$
being nonempty, where $\nu$ is the outward unit normal vector on $\partial \Omega$, and $\xi\gg 0$ means that $\xi(t,x)>0$ for all $(t,x)\in [0, \tau] \times \Omega$ and $\frac{\partial \xi}{\partial \nu} (t,x)<0$ for all $(t,x)\in [0, \tau] \times \partial \Omega$, normally called as a strongly-positive function in $[0,\tau] \times \Omega$.

Let $M^*=1+\max_{[0,\tau]\times \overline \Omega} |a(t,x)|+\ln (\max\{1/z,1\})$. For any  $\xi(t,x)\in W$ given, we first conclude that the linear problem
\begin{eqnarray}
\left\{
\begin{array}{lll}
\phi_t-d\Delta \phi=(-M^*+a(t,x)-yb(t,x))\phi+\xi,& t\in(0,\tau],\,\,x \in \Omega,  \\[1mm]
\phi(t,x)=0,\  & t\in [0, \tau],\,\,x\in \partial\Omega,\\[1mm]
\phi(0,x)=z\phi(\tau,x), & x\in \overline \Omega
\end{array} \right.
\label{3.22}
\end{eqnarray}
admits a unique solution $\phi \in {C}^{(1+\alpha)/2, 1+\alpha}([0, \tau]\times \overline \Omega)\bigcap W$ for some $0<\alpha<1$.

Next, define $\mathcal{A} \xi = \phi$. Owing to the embedding ${C}^{(1+\alpha)/2,1+\alpha}\hookrightarrow {C}^{0,1}$ be compact, $\mathcal{A}$ is a linear compact operator. Moreover, $\mathcal{A}$ is strongly positive with respect to
$W$ according to the strong maximum principle and Hopf's boundary lemma. Thus, by the use of Krein-Rutman theorem, we obtain that there exist a unique $\sigma_1 := r(\mathcal{A}) > 0$ and a function $\phi\in \textrm{Int} (W^+)$ such that $\mathcal{A}\phi = \sigma_1 \phi$, and therefore
$$\mu_1 := 1/\sigma_1-M^*$$
is the principal eigenvalue of \eqref{3.02} and its corresponding
eigenfunction $\phi$ satisfies $\phi(t,x)>0$ in $[0, \tau]\times \Omega$ and $\frac{\partial \xi}{\partial \nu} (t,x)<0$ for $(t,x)\in [0, \tau] \times \partial \Omega$, that is, $\phi\gg 0$.

Coming back to the problem \eqref{a013} with impulse, we now have the existence of the principal eigenvalue $\mu_1$, whose monotonicity with respect to $d$ and $y$ follows from similar arguments as done in \cite{AL2017, LG2020}.

\begin{lem}\label{l21} Problem \eqref{a013} admits a principal eigenvalue $\mu_1:=\mu_1(d, a(t,x), yb(t,x), z)$ with its corresponding
eigenfunction $\phi(t,x) \gg 0$ in $[0,\tau] \times \Omega$. Moreover, $\mu_1(d, a(t,x), yb(t,x), z)$ is increasing in $d>0$ and $y\geq 0$.
\label{abd}
\end{lem}

By the periodicity of the terms in problem \eqref{a013}, let us use indistinctly $\mu_1$ and $\phi$ as the principal eigenvalue and the corresponding eigenfunction, respectively, of the problem
\begin{eqnarray}
\left\{
\begin{aligned}
&\phi_t-d \Delta \phi-a(t,x)\phi+ yb(t, x)\phi=\mu \phi, && t\in((n\tau)^+,(n+1)\tau],\,\,x \in \Omega,  \\
&\phi(t,x)=0, && t\in [0, \tau],\,\,x \in \partial \Omega,\\
&\phi((n\tau),x)=\phi((n+1)\tau,x), && x\in \overline \Omega,\ n=0,1,2,\cdots,\\
&\phi((n\tau)^+,x)=z\phi(n\tau,x),&& x\in \overline \Omega,\ n=0,1,2,\cdots,
\end{aligned} \right.
\label{a01311}
\end{eqnarray}
for any $\tau,z>0$ given.

Besides the eigenvalue problem \eqref{a013} (or \eqref{a01311}), the next one will play an important tool in many of our proofs. Let the periodic $\lambda$-eigenvalue problem
\begin{eqnarray}
\left\{
\begin{aligned}
&-\psi_t-d \Delta \psi-a(t,x)\psi+ yb(t, x)\psi=\lambda \psi, && t\in(0^+, \tau],\,\,x \in \Omega,  \\
&\psi(t,x)=0, && t\in [0, \tau],\,\,x \in \partial \Omega,\\
&\psi(0,x)=\psi(\tau, x), && x\in \overline \Omega,\\
&\psi(0^+, x)=\frac{1}{z}\psi(0, x)),&& x\in \overline \Omega.
\end{aligned} \right.
\label{e28}
\end{eqnarray}

So, we have the below result.
\begin{lem}\label{l23} Problem \eqref{e28} admits a principal eigenvalue $\lambda_1 := \lambda_1(d, a(t,x), yb(t,x), z)$. Its corresponding
eigenfunction is $\psi$ satisfying $\psi(t,x) \gg 0$ in $[0,\tau] \times \Omega$. Moreover, $\lambda_1=\mu_1$, where $\mu_1$ is the principal eigenvalue of problem \eqref{a013}.
\label{abd1}
\end{lem}
\bpf
The existence of the principal eigenvalue $\lambda_1 := \lambda_1(d, a(t,x), yb(t,x), z)$ follows the same arguments as done in the proof of Lemma  \ref{l21} to prove the existence of $\mu_1$. Below let us prove that $\lambda_1=\mu_1$. Let $(\mu_1, \phi)$ and $(\lambda_1, \psi)$ be the pairs of principal eigenvalues and eigenfunctions of the eigenvalues problems \eqref{a013} and \eqref{e28}, respectively.

So, by multiplying the first equation of problem  $\eqref{a013}$ by $\psi$, we obtain
$$\psi \phi_{t}-d \psi\Delta \phi-a(t,x)\phi\psi+ y b(t,x)\phi\psi=\mu_1 \phi\psi,$$
while the product of $\phi$ by the first equation of the problem $\eqref{e28}$ leads us to the equation
$$-\phi \psi_{t}-d \phi \Delta \psi-a(t,x)\phi\psi+ y b(t,x)\psi\phi=\lambda_1 \phi \psi.$$

Now, by subtracting the last equation from the former, one yields
$$(\psi\phi)_t-d( \psi \Delta \phi - \phi \Delta \psi)=(\mu_1 - \lambda_1) \psi\phi,$$
whence follows by integration in $ x \in \Omega$, $t \in (0^+,\tau]$, and the other conditions in problems \eqref{a013} and  \eqref{e28}, that
$$(\mu_1 - \lambda_1)\int^\tau_{0^+}\int_\Omega \phi \psi =0 $$
leading to  $\mu_1 = \lambda_1$. This ends the proof.
\epf

Due to the importance that the variable $y\geq 0$ and $z>0$ will play on our arguments, let us denote by
$$\Sigma (y,z)=\mu_1(d, a(t,x), yb(t,x), z)$$
the principal eigenvalue of problem \eqref{a013} for fixed $d, a(t,x)$ and $b(t,x)$, and by
 $$\Sigma (\infty,z)=\lim_{y\to\infty} \Sigma (y,z) \leq \infty,$$
 for any $z>0$ fixed. The above limit is well defined due to the fact that  $\Sigma (y,z)$ be increasing with respect to $y\geq 0$ for any $z>0$ fixed.
 The following results brings up more properties on $\Sigma (y,z)$.
\begin{thm}\label{t23} Let $z>0$.
\begin{enumerate}
    \item[$(i)$] if $a(t,x)=a(t)$ for $t\in [0, \tau]$ and $x\in \overline \Omega$, then
    $$\Sigma (0,z)=d\lambda_0-\frac 1\tau \int^\tau_0 a(s)ds-\frac{\ln z}{\tau},$$  where $\lambda_0>0$ is the principal eigenvalue  of $-\Delta$ in $\Omega$ under Dirichlet boundary condition;
     \item[$(ii)$] if $b(t,x)>0$ in $[0,\tau]\times\overline\Omega$, then
     $$\Sigma (\infty,z):=\lim_{y\to\infty} \Sigma(y,z)=+\infty, \ \mbox{for any }z>0 \ \mbox{fixed};$$
     \item[$(iii)$] if $b(t,x)\equiv 0$ in $[0,\tau]\times \overline\Omega$, then
     $$ \mu_1(d, a_M (t), yb(t,x), z)\leq \Sigma (0,z) =\Sigma (\infty,z)\leq \mu_1(d, a_m (t), yb(t,x), z),$$
     where
     $$a_m (t) = \min_{x\in \overline \Omega}a(t,x)) \ \textrm{and }\, a_M (t) = \max_{x\in \overline \Omega}a(t,x); $$
      \item[$(iv)$] if $b(t,x)\not\equiv 0$ in $[0,\tau]\times\overline\Omega$, then $\Sigma(y,z)$ is a continuous and strictly increasing function of $y \geq 0$ for each $z>0$;
      \item[$(v)$]  Assume that $b(t,x)>0$ on $[0, \tau]\times \partial \Omega$. Then, for each $z>0$, $\Sigma (\infty,z)<+\infty$ if, and only if, there exists a continuous map $\kappa:\, [0, \tau]\to \Omega$ such that $(t,\kappa(t))\in \,\textrm{int}\, b^{-1}(0)$ for all $t\in [0, \tau]$.
\end{enumerate}
\end{thm}
\bpf
Item $(i)$. Under the assumption $a(t,x)=a(t)$ for $t\in [0, \tau]$ and $x\in \overline \Omega$, the eigenvalue problem \eqref{a013}, for $y=0$, becomes
\begin{eqnarray}
\label{e1}
\left\{
\begin{aligned}
&\phi_t-d \Delta \phi-a(t)\phi=\Sigma (0,z) \phi, && t\in(0^+, \tau],\,\,x \in \Omega,  \\
&\phi(t,x)=0, && 0<t\leq \tau,\,\,x \in \partial \Omega,\\
&\phi(0,x)=\phi(\tau, x), && x\in \overline \Omega,\\
&\phi(0^+, x)=z\phi(0, x),&& x\in \overline \Omega.
\end{aligned} \right.
\end{eqnarray}

By taking
\begin{eqnarray}
f(t)=\left\{
\begin{array}{lll}
1, \; &\, t=0, \\[1mm]
z,\; &\,t=0^+,\\[1mm]
z e^{(-d\lambda_0+\frac 1\tau \int^\tau_0 a(s)ds+\Sigma(0,z))t},\;&\,t\in(0^+, \tau]
\end{array} \right.
\label{c101}
\end{eqnarray}
and defining $\phi(t,x)=f(t)\psi(x)$, it is straightforward to show that the principal eigenvalue $\Sigma (0,z)$ is explicitly expressed as
$$\
\Sigma(0,z)=d\lambda_0-\frac 1\tau \int^\tau_0 a(s)ds-\frac{\ln z}{\tau},
$$
where $\lambda_0>0$ is the principal eigenvalue  of $-\Delta$ in $\Omega$ under Dirichlet boundary condition and $\psi(x)>0$ in  $\Omega$ is its corresponding eigenfunction.

 The item $(ii)$ is a consequence of \cite[Theorem 6.2 (b)]{AL2017} due to the fact that
$$\int^\tau_0 \min_{x\in  \overline \Omega} b(s,x)ds>0.$$

 About the item $(iii)$. Since we are assuming that $b(t,x)\equiv 0$ in $[0,\tau] \times \overline\Omega$, problem \eqref{a013} becomes
 \begin{eqnarray}
\left\{
\begin{aligned}
&\phi_t-d \Delta \phi-a(t,x)\phi=\Sigma (y,z) \phi, && t\in(0^+, \tau],\,\,x \in \Omega,  \\
&\phi(t,x)=0, && t\in [0, \tau],\,\,x \in \partial \Omega,\\
&\phi(0,x)=\phi(\tau, x), && x\in \overline \Omega,\\
&\phi(0^+, x)=z\phi(0, x),&& x\in \overline \Omega
\end{aligned} \right.
\label{a130}
\end{eqnarray}
so that
$$\Sigma (0,z)=\Sigma (\infty,z)=\mu_1(d, a(t,x), 0, z)\leq  \mu_1(d, a_m (t), yb(t,x), z),$$
for each $z>0$, due to the fact that $\mu_1$ be increasing with respect to $a(t,x)$, and the item  $(i)$ just proved. Similarly, we prove the estimate from below.

 About the item $(iv)$. Under the assumption $b(t,x)\not\equiv 0$ in $[0,\tau]\times\overline \Omega$, we can take advantage of the ideas of the proofs of \cite[Corollary 2.3 and Lemma 2.2]{AL2017} to show the $\Sigma(y,z)$ is continuous and the strict increasing with respect to $y\geq 0$.

 Finally, by considering the equivalent eigenvalue problem \eqref{3.02}, instead of \eqref{a013}, the proof of $(v)$ follows, with minors adjustments, from the proof of the main result in \cite{LG2020}, after using a tricky of a novel generalized version of the parabolic maximum principle on general domains. This ends the proof.
\epf

\begin{thm}\label{t24} For any $y\geq 0$ and $z_1, z_2>0$, we have
\begin{equation}\label{ia28}
\Sigma (y, z_1)-\Sigma (y, z_2)=\frac {\ln(z_2/z_1)}\tau.
\end{equation}
In particular, for any $y\geq 0$, one has:
\begin{enumerate}
    \item[$(i)$] $\Sigma (y, z)$ is continuous and strictly decreasing with respect to $z>0$,
     \item[$(ii)$]  $\lim_{z\to 0^+} \Sigma(y,z)=+\infty$ and $\lim_{z\to\infty} \Sigma(y,z)=-\infty$,
\end{enumerate}
\end{thm}
\bpf First, we point out that the proof of the above items $(i)$ and $(ii)$ are an immediate consequence of the equality \eqref{ia28}. To prove  \eqref{ia28}, let $(\Sigma (y, z_1),\phi)$ be the eigen-pair of the eigenvalue problem \eqref{a013}, that is,
\begin{eqnarray*}
\left\{
\begin{aligned}
&\phi_t-d \Delta \phi - a(t,x) \phi+yb(t,x)\phi=\Sigma (y, z_1)\phi, && t\in(0^+, \tau],\,\,x \in \Omega,  \\
&\phi(t,x)=0, && 0<t\leq \tau,\,\,x \in \partial \Omega,\\
&\phi(0,x)=\phi(\tau, x), && x\in \overline \Omega,\\
&\phi(0^+, x)=z_1\phi(0, x),&& x\in \overline \Omega.
\end{aligned} \right.
\label{a021}
\end{eqnarray*}
By taking
$$\psi(t,x)=e^{\frac {\ln(z_1/z_2)}{\tau}t}\phi(t,x),\  x \in \Omega, \ t\in(0^+, \tau],\ \mbox{and } \psi(0,x)=\psi(\tau,x)=z_1/z_2 \phi(\tau,x),$$
we obtain that $\psi$ satisfies
$$
\left\{
\begin{aligned}
&\psi_t-d \Delta \psi- a(t,x) \psi+yb(t,x)\psi=\left(\Sigma (y, z_1)+\frac {\ln(z_1/z_2)}{\tau}\right)\psi, && t\in(0^+, \tau],\,\,x \in \Omega,  \\
&\psi(t,x)=0, && 0<t\leq \tau,\,\,x \in \partial \Omega,\\
&\psi(0,x)=\psi(\tau, x), && x\in \overline \Omega,\\
&\psi(0^+, x)=\phi(0^+,x)=z_1\phi(0,x)=z_1\phi(\tau,x)=z_2\psi(0, x),&& x\in \overline \Omega,
\end{aligned} \right.
\label{a022}
$$
whence follows by the uniqueness of the principal eigenvalue of the above eigenvalue problem that the equality \eqref{ia28} holds.
This ends the proof.
\epf

About $\Sigma (\infty, z)$, for $z>0$, we obtain form the above theorem the next result.
\begin{thm}\label{t25} Let $z_0>0$ given. Then one has:
\begin{enumerate}
    \item[$i)$] either $\Sigma (\infty, z_0) = \infty $  if, and only if, $\Sigma (\infty, z) = \infty $ for all $z>0$,
    \item[$ii)$] or  $\Sigma (\infty, z_0) < \infty $  if, and only if, $\Sigma (\infty, z) < \infty $ for all $z>0$. In such case, we have:
    \begin{enumerate}
    \item[$(ii_1)$] $\Sigma (\infty, z)$ is continuous and strictly decreasing with respect to $z>0$,
     \item[$(ii_2)$]  $\lim_{z\to 0^+}\Sigma(\infty, z)=+\infty$ and $\lim_{z\to\infty} \Sigma(\infty, z)=-\infty$.
\end{enumerate}
\end{enumerate}
\end{thm}
\bpf It is immediate from the equality \eqref{ia28} by passing to the limit as $y\to \infty$.
\epf

\begin{rmk}
    After Theorem $\ref{t23}$-$(v)$ and Theorem $\ref{t25}$, we deduce that the subset $b^{-1}(0)$ will work as a periodic-parabolic protection zone for the species if, and only if, $\Sigma(\infty, c_0)<\infty$ for some $c_0>0$, that is, that size of the impulse $c>0$ does not change the status of the protection zone if it occurs for some $c_0>0$.
\end{rmk}

\section{\bf About a impulsive periodic problem associated to Problem \eqref{a01} }

It is well known from \cite{MGL} that problem \eqref{a01} admits a unique global classical solution $u(t,x)\in C^{1+\theta/2, 2+\theta}((n\tau, (n+1)\tau]\times \Omega)\bigcup C((n\tau, (n+1)\tau])\times\overline \Omega)$ $(n=0,1,2,\cdots)$ for any nonnegative and continuous initial value $u_0$ that vanishes on the boundary $\partial \Omega$, and, as in \cite{DLG2018}, the dynamic of problem \eqref{a01} is related to its corresponding periodic problem
\begin{eqnarray}
\left\{
\begin{aligned}
&U_t=d \Delta U+ a(t,x) U-b(t,x)U^p, && t\in(0^+, \tau],\,\,x \in \Omega,  \\
&U(t,x)=0, && 0<t\leq \tau,\,\,x \in \partial \Omega,\\
&U(0,x)=U(\tau, x), && x\in \overline \Omega,\\
&U(0^+, x)=cU(0, x),&& x\in \overline \Omega.
\end{aligned} \right.
\label{a012}
\end{eqnarray}

So, we establish the next theorem.
\begin{thm}\label{ap} One has.
\begin{enumerate}
    \item[$(i)$] If either $b(t,x)\not\equiv 0$ in $[0,\tau]\times \overline\Omega$  and $\Sigma (0,c)\geq 0$ or $b(t,x)\equiv 0$  in $[0,\tau]\times \overline\Omega$ and $\Sigma (0,c) \neq 0$, then problem \eqref{a012} has no positive solution.
 \item[$(ii)$] If either $b(t,x)\not\equiv 0$  in $[0,\tau]\times \overline\Omega$ and $\Sigma (\infty, c)\leq 0$ or $b(t,x)\equiv 0$ in $[0,\tau]\times \overline\Omega$ and $\Sigma (\infty, c) \neq 0$, then problem \eqref{a012} has no positive solution.
     \item[$(iii)$] If $\Sigma(0,c)<0<\Sigma(\infty,c)$ $(b(t,x)\not\equiv 0$ in this case$)$,  then the periodic problem \eqref{a012} admits a unique positive periodic solution, say $U(t,x)$.
\end{enumerate}
\end{thm}
\bpf
For the item $(i)$. It follows from Lemma \ref{l23} that  $\Sigma (0,c)$ is the principal eigenvalue of  problem \eqref{e28} with $y=0$, that is,
\begin{eqnarray}
\left\{
\begin{aligned}
&-\psi_t-d \Delta \psi-a(t,x)\psi=\Sigma (0,c) \psi, && t\in(0^+, \tau],\,\,x \in \Omega,  \\
&\psi(t,x)=0, && 0<t\leq \tau,\,\,x \in \partial \Omega,\\
&\psi(0,x)=\psi(\tau, x), && x\in \overline \Omega,\\
&\psi(0^+, x)=\frac{1}{c}\psi(0, x)),&& x\in \overline \Omega,
\end{aligned} \right.
\label{a105}
\end{eqnarray}
where $\psi \gg 0$ denotes the eingenfunction associate to $\Sigma (0,c)$.

Assume, by contradiction, that problem \eqref{a012} admited a positive solution $U(t,x)$. Following the same strategy as done in the proof of Lemma \ref{l23}, we would obtain
$$(U\psi)_t=d(\Delta U \psi-\Delta \psi U)-\Sigma(0,c) U \psi-b(t,x) U^p \psi,$$
that leads, after integrating in $x\in \Omega$ and $t$ from $0^+$ to $\tau$ together with the use of the other conditions in \eqref{a012} and  \eqref{a105}, to
\begin{align*}
   -&\Sigma(0,c)\int^\tau_0\int_\Omega U \psi dxdt-\int^\tau_0\int_\Omega b(t,x) U^p \psi dxdt  \\
=&\int_\Omega [U(\tau,x) \psi(\tau,x)-U(0^+,x)\psi(0^+,x)]dx\\
=&\int_\Omega [U(0,x) \psi(0,x)-c U(0,x)\frac{1}{c}\psi(0, x)]dx\\
= &\ 0.
\end{align*}

So, we must have $\Sigma(0,c)<0$ when $b(t,x)\not\equiv 0$ in $[0,\tau]\times \overline\Omega$, and  $\Sigma(0,c)=0$ when $b(t,x)\equiv 0$ in $[0,\tau]\times \overline\Omega$. Both situations contradict our assumptions.

 Proof of item $(ii)$. Let us proceed by contradiction again. Suppose that problem \eqref{a012} admits a positive solution $U(t,x)$. Then $U(t,x)$ is bounded from above by some positive constant $M^{1/(p-1)}$.

Since $\Sigma(y,c)$ is the principal eigenvalue of problem \eqref{e28}, see Lemma \ref{l23}, that is,
\begin{eqnarray*}
\left\{
\begin{aligned}
&-\psi_t- d \Delta \psi- a(t,x)\psi+y b(t,x)\psi=\Sigma(y,c)\psi, && t\in(0^+, \tau],\,\,x \in \Omega,  \\
&\psi(t,x)=0, && 0<t\leq \tau,\,\,x \in \partial \Omega,\\
&\psi (0,x)=\psi(\tau, x), && x\in \overline \Omega,\\
&\psi(0^+, x)=\frac 1{c}\psi(0, x),&& x\in \overline \Omega,
\end{aligned} \right.
\end{eqnarray*}
where $\psi \gg 0$ is the first eigenfunction associate to $\Sigma(y,c)$,  we obtain, by following the same arguments as done in the proof of Lemma \ref{l23}, that
$$(U\psi)_t-d(\psi \Delta U -U\Delta \psi ) = y b(t,x) U \psi-b(t,x) U^p \psi -\Sigma(y,g'(0)) U \psi$$
holds, whence follows by integration in $x\in \Omega$, for $t$ from $0^+$ to $\tau$, that
\begin{align*}
    -&\Sigma(y,c)\int^\tau_0\int_\Omega U \psi dxdt + \int^\tau_0\int_\Omega b(t,x) U \psi \left(y - U^{p-1} \right) dxdt\\
    =& \int_\Omega [U(\tau,x) \psi (\tau,x)-U(0^+,x)\psi(0^+,x)]dx\\
=&\int_\Omega [U (0,x)\psi(0,x)-cU(0,x)\frac{1}{c}\psi(0, x)]dx\\
= &\ 0,
\end{align*}
holds for any $y \geq 0$. So, when $b(t,x)\not\equiv 0$  in $[0,\tau]\times \overline\Omega$, we have
$$\Sigma(y,c) > 0\ \mbox{for any } y> M,$$
whence follows, together with the fact that $\Sigma(y,c) > 0$ is strict increasing with respect to $y \geq 0$,   that
$$\Sigma(\infty,c) = \lim_{y \to \infty} \Sigma(y,c) >0,$$
which is  a contradiction to our assumption $\Sigma(\infty,c)\leq 0$. If  $b(t,x)\equiv 0$ in $[0,\tau]\times \overline\Omega$, then $ \Sigma(\infty,c) = \Sigma(0,c) \neq 0$, as proved in the item $(i)$.

 Proof of item $(iii)$. It suffices to construct an upper solution and a lower solution  for the problem \eqref{a012}. Since $\Sigma(0,c)<0<\Sigma(\infty,c)$, we have $b(t,x)\not\equiv 0$ in $[0,\tau]\times \overline\Omega$, and there exist
$\delta>0$ and $M>0$ such that
$$\Sigma(0,c)<\Sigma(\delta,c)<0<\Sigma(M,c)<\Sigma(\infty,c).$$

Let us first construct an upper solution $\overline{U}$ for the periodic problem \eqref{a012} inspired on ideas from \cite[Proposition 3.2]{DLG2018}. Since $\Sigma(M,c)>0$, we are able to find an  $\varepsilon>0$ such that
$$0<\Sigma_\varepsilon (M,c):=\mu_1[d,a(t,x), Mb(t,x)\chi_{\Omega_\varepsilon}(x),c],$$
where
$$\Omega_\varepsilon:=\{x\in\Omega:\,\textrm{dist}\,(x,\partial \Omega)>\varepsilon\},\ \chi_{\Omega_\varepsilon}(x)=1\, \textrm{for}\, x\in \Omega_\varepsilon \ \textrm{and}\, \chi_{\Omega_\varepsilon}(x)=0\, \textrm{for}\, x\in \Omega/ \Omega_\varepsilon.$$

Let $\varphi \gg 0$ be the eigenfunction corresponding to the principal eigenvalue $\Sigma_\varepsilon (M,c)$ of problem \eqref{a013} with $\max_{(t,x)\in [0,\tau]\times \overline \Omega}\varphi(t,x)= 1$ (here, we are considering problem  \eqref{a013} with $ b(t,x)\chi_{\Omega_\varepsilon}$ in the place of $b(t,x)$). That is,
\begin{eqnarray*}
\left\{
\begin{aligned}
&\varphi_t-d\Delta\varphi-a(t,x)\varphi+M b(t,x)\chi_{\Omega_\varepsilon}(x)\varphi=\Sigma_\varepsilon (M,c)\varphi, && t\in(0^+, \tau],\,\,x \in \Omega,  \\
&\varphi(t,x)=0, && t\in [0, \tau],\,\,x \in \partial \Omega,\\
&\varphi(0,x)=\varphi(\tau, x), && x\in \overline \Omega,\\
&\varphi(0^+, x)=c\varphi(0, x)),&& x\in \overline \Omega.
\end{aligned} \right.
\label{a0131}
\end{eqnarray*}

By defining
\begin{equation}\label{d32}
    \overline{U} =K\varphi,
\end{equation}
 we obtain
\begin{align*}
&\overline{U}_t-d\Delta \overline{U}-a(t,x)\overline{U}+b(t,x)\overline{U}^p\\
=&\ \overline{U}[\Sigma_\varepsilon (M,c)+(-M\chi_{\Omega_\varepsilon}(x) +K\varphi)b(t,x)]\\
>& \ \overline{U}[-M\chi_{\Omega_\varepsilon}(x) +K\varphi]b(t,x)\\
\geq&\ 0
\end{align*}
for any $K\geq M/\min_{[0,\tau]\times \overline \Omega_\varepsilon}\, [\varphi(t,x)]$ given.

Besides this, it is immediate to verify that $\overline{U}(t,x)=0$ for $t\in [0, \tau],\,\,x \in \partial \Omega$, $\overline{U}(0,x)=\overline{U}(\tau, x), \ x\in \overline \Omega$, and
$\overline{U}(0^+,x)-c\overline{U}(0,x)=0$ for $x\in \overline \Omega$ whence follows that
 $\overline{U}$ is an upper solution for the problem \eqref{a012}.

In what follows, let us build a lower solution, say $\underline{U}$. To do this, define
\begin{eqnarray}
\underline{U}(t,x)=
\begin{cases}
\displaystyle \varepsilon{\phi}(0,x), &t=0,\,\,x\in\overline\Omega, \\[3mm]
\displaystyle \varepsilon{\frac{\rho}{c}}\phi(0^+,x), &t=0^+,\,\,x\in\overline\Omega,\\[3mm]
\displaystyle \varepsilon{\frac{\rho}{c}}e^{-\Sigma(\delta,c) t}\phi(t,x),&t\in(0^+,\tau],\,\,x\in\overline\Omega,
\end{cases}
\label{4.52}
\end{eqnarray}
where $\varepsilon>0$ is a sufficiently small constant, and $\phi \gg 0$ is the eigenfunction associated to the principal eigenvalue $\Sigma(\delta,c)$ of the problem \eqref{a013} with $\max_{(t,x)\in [0,\tau]\times \overline \Omega}\phi(t,x)= 1$, that is,
\begin{eqnarray*}
\left\{
\begin{aligned}
&\phi_t-d\Delta\phi=a(t,x)\phi-\delta b(t,x)\phi+\Sigma(\delta,c)\phi, && t\in(0^+, \tau],\,\,x \in \Omega,  \\
&\phi(t,x)=0, && t\in [0, \tau],\,\,x \in \partial \Omega,\\
&\phi(0,x)=\phi(\tau, x), && x\in \overline \Omega,\\
&\phi(0^+, x)=c\phi(0, x)),&& x\in \overline \Omega.
\end{aligned} \right.
\label{a013y}
\end{eqnarray*}

Let us choose the parameters. First, we point out that $\underline{U}(0,x)=\underline{U}(\tau,x)$ holds if and only if  $\rho=ce^{\Sigma(\delta,c)\tau}$. By setting such $\rho>0$ and noting that $\rho=ce^{\Sigma(\delta,c)\tau}<c$, due to the fact that $\Sigma(\delta,c)<0$, we obtain
that
\begin{align*}
&\underline{U}_t-d\Delta \underline{U}-a(t,x)\underline{U}+b(t,x)\underline{U}^p\\
=&\ \underline{U}[-\delta +\varepsilon\frac{\rho}{c}e^{-\Sigma(\delta,c)t}\phi)b(t,x)]\\
\leq&\ \underline{U}(-\delta +\varepsilon e^{-\Sigma(\delta,c)\tau})b(t,x)\\
\leq &\ 0
\end{align*}
for any
$$0<\varepsilon \leq \delta e^{\Sigma(\delta,c)\tau}$$
given. It is an immediate consequence of the definition of $\underline{U}$ that $\underline{U}(t,x)=0, \ t>0,x\in\partial\Omega$, and $\underline{U}(0^+,x)=c\underline{U}(0,x),\  x\in\Omega$ so that $\underline{U}$ is a lower solution to problem \eqref{a012}.

By taking  $\varepsilon>0$ small enough, if it necessary, we obtain $\underline{U}\leq \overline{U}$. Then, based on standard upper and lower solutions technique \cite[Theorem 3.4]{MLP2021}, we have the existence of a $\tau-$periodic solution for problem \eqref{a012}.

For the uniqueness of $\tau-$periodic solution, suppose that $U_1$ and $U_2$ are two solutions of problem \eqref{a012}, and define
$$
S=\{s\in{[0,1]},sU_1\leq{U_2},\,t\in{[0,\tau]},\,x\in\overline \Omega\}.
$$
By using the fact that $f(v)/v$ is non-increasing with respect to $v$ in $(0, \max\limits_{[0,T]\times{\overline\Omega}}U_2]$, where $f(v):=a(t,x) v-b(t,x)v^p$, we are able to prove, in a similar way as was done in  \cite[Theorem 3.4]{MLP2021}, that that $1\in {S}$ so that  $U_1\leq U_2$. Similarly, we have $U_2\leq U_1$ that leads to the uniqueness of solutions to the problem \eqref{a012}. This ends the proof.
\epf

Again, by the periodicity of the terms $a(t,x)$ and $b(t,x)$ and the eigenvalue properties, we will naturally extend the conclusions of Theorem \ref{ap} to the periodic-parabolic $\lambda$-problem
\begin{eqnarray}
\left\{
\begin{aligned}
&U_t-d \Delta U= a(t,x) U-b(t,x)U^p + \lambda U, && t\in((n\tau)^+,(n+1)\tau],\,\,x \in \Omega,  \\
&U(t,x)=0, && t\in [0, \tau],\,\,x \in \partial \Omega,\\
&U((n\tau),x)=U((n+1)\tau,x), && x\in \overline \Omega,\ n=0,1,2,\cdots,\\
&U((n\tau)^+,x)=c U(n\tau,x),&& x\in \overline \Omega,\ n=0,1,2,\cdots.
\end{aligned} \right.
\label{p351}
\end{eqnarray}
That is.
\begin{cor}\label{ap2} One has.
\begin{enumerate}
    \item[$(i)$] If either $b(t,x)\not\equiv 0$ in $[0,\tau]\times \overline\Omega$  and $\Sigma (0,c)\geq \lambda $ or $b(t,x)\equiv 0$  in $[0,\tau]\times \overline\Omega$ and $\Sigma (0,c) \neq \lambda$, then problem \eqref{p351} has no positive solution.
 \item[$(ii)$] If either $b(t,x)\not\equiv 0$  in $[0,\tau]\times \overline\Omega$ and $\Sigma (\infty, c)\leq \lambda$ or $b(t,x)\equiv 0$ in $[0,\tau]\times \overline\Omega$ and $\Sigma (\infty, c) \neq \lambda$, then problem \eqref{p351} has no positive solution.
     \item[$(iii)$] If $\Sigma(0,c)<\lambda<\Sigma(\infty,c)$, then the periodic problem \eqref{p351} admits a unique positive periodic solution, say $U_\lambda(t,x)$.
\end{enumerate}
\end{cor}

About the behavior at the endpoints of the $\lambda$-interval for the existence, we have the next result.
\begin{thm} If $b(t,x)\not\equiv 0$ in $[0,\tau]\times \overline\Omega$, then the solution $U_\lambda$ is strictly increasing in $\lambda \in (\Sigma(0,c),\, \Sigma(\infty,c))$ in the sense that  $U_{\lambda_2}\gg U_{\lambda_1}$ if $\lambda_2>\lambda_1$. Moreover,
\begin{eqnarray}\lim_{\lambda \downarrow\Sigma(0,c)}\, \Vert U_\lambda\Vert_{C([0,\tau] \times \overline{\Omega})} =0\label{ma1}
\end{eqnarray}
and if $\Sigma(\infty,c)<\infty$, then
\begin{eqnarray}\lim_{\lambda \uparrow\Sigma(\infty,c)}\, \Vert U_\lambda\Vert_{C([0,\tau] \times \overline{\Omega})} =\infty.\label{ma2}
\end{eqnarray}
\end{thm}
\bpf First, we note that is a consequence of Theorem \ref{t23}-$(iv)$ that  $b(t,x)\not\equiv 0$ in $[0,\tau]\times \overline\Omega$ leads to $\Sigma(0,c) < \Sigma(\infty,c)$.
Assume $\lambda_2>\lambda_1$. We first claim that $U_{\lambda_2}\geq U_{\lambda_1}$ in $[0,\tau]\times\overline \Omega$,
where $U_{\lambda_i}$ is the solution of problem \eqref{p351} with $\lambda$ replaced with $\lambda_i$ for $i=1,2$.
In fact, $U_{\lambda_1}$ is a lower solution for problem \eqref{p351} with $\lambda=\lambda_2$. As done in the proof of Theorem \ref{p351}-$(iii)$, we are able to build a solution $\overline{U} $ satisfying $U_{\lambda_1} \leq \overline{U}$. So, the claim is derived by a iteration process, by using the lower solution as the initial iteration, and the uniqueness of solution for problem \eqref{p351}.
%{\color{red} I stopped here}

Next let us prove that $U_{\lambda_2}\gg U_{\lambda_1}$.
To circumvent the difficulties induced by the impulse in the parabolic-periodic problem \eqref{p351}, we consider the following equivalent problem
\begin{eqnarray}
\left\{
\begin{array}{lll}
(V_\lambda)_t=d \Delta V_\lambda+a(t,x)V_\lambda-b(t,x)V_\lambda^p+\lambda V_\lambda,& t\in(0,\tau],\,\,x \in \Omega,  \\[1mm]
V_\lambda(t,x)=0, & t\in [0, \tau],\,\,x\in \partial\Omega,\\[1mm]
V_\lambda(0,x)=c V_\lambda(\tau,x), & x\in \overline \Omega.
\end{array} \right.
\label{3.42}
\end{eqnarray}
This equivalence is reached by taking
\begin{equation}\label{e38}
U_\lambda(t,x)=V_\lambda(t,x)\ \mbox{for } t\in (0,\tau],\ x\in \overline \Omega
\end{equation}
and
\begin{equation}\label{e39}
U_\lambda(0^+,x)=V_\lambda(0,x)\ \mbox{for } x\in \overline \Omega,\ \mbox{and }  U_\lambda(0,x)=V_\lambda(\tau, x) \ \mbox{for } x\in \overline \Omega.
\end{equation}
So, the proof that $U_{\lambda_2}\gg U_{\lambda_1}$ holds is obtained by proving that $V_{\lambda_2}\gg V_{\lambda_1}$, where $V_{\lambda_i}$ is the solution of the problem \eqref{3.42} with $\lambda$ replaced with $\lambda_i$ for $i=1,2$.

 Let $W=V_{\lambda_2}-V_{\lambda_1}$. Then due to  $U_{\lambda_2}\geq U_{\lambda_1}$, we have that $W \geq0$ and satifies
 \begin{eqnarray}
\left\{
\begin{array}{lll}
W_t=d \Delta W+a W+b\xi(t,x) W+(\lambda_2-\lambda_1) V_{\lambda_2}+\lambda_1W,& t\in(0,\tau],\,\,x \in \Omega,  \\[1mm]
W(t,x)=0, & t\in [0, \tau],\,\,x\in \partial\Omega,\\[1mm]
W(0,x)=cV_{\lambda_2}(\tau,x)-c V_{\lambda_1}(\tau,x)=c W(\tau,x), & x\in \overline \Omega,
\end{array} \right.
\label{3.43}
\end{eqnarray}
where
$$
\xi(t,x) =p \int_0^1\left((1-s) V_{\lambda_2}(t,x) + s V_{\lambda_1}(t,x)\right)^{p-1} ds.
$$

If there was a $(t_0, x_0)\in [0,\tau]\times \Omega$ such that $W(t_0,x_0)=0 = \min_{[0,\tau]\times \overline\Omega}
W$, then we could choose $t_0\in (0,\tau]$, and $W\equiv 0$ in $[0,t_0]\times \overline \Omega$ by the strong maximum principle.
Hence $W(\tau,x)\equiv W(0,x)\equiv 0$ in $\overline \Omega$ due to $W(0,x)=c W(\tau,x)$ in $\Omega$,
which turns $W\equiv 0$ in $[0,\tau]\times \overline\Omega$. This indicates that $W_t=d \Delta W+a W-b\xi(t,x) W+\lambda_1W\equiv 0$
 in $(0,\tau]\times \Omega$, that leads  to a contradiction with the first equation of \eqref{3.43}, since $(\lambda_2-\lambda_1)V_{\lambda_2}>0$ in $(0, \tau]\times \Omega$.

For the remaining two limits, we only prove that $\lim_{\lambda \uparrow\Sigma(\infty,c)}\,  \Vert U_\lambda\Vert_{C([0,\tau] \times \overline{\Omega})} =\infty$, because the proof of $\lim_{\lambda \downarrow\Sigma(0,c)}=0$ is similar. Equivalently, it suffices to prove
$\lim_{\lambda \uparrow\Sigma(\infty,c)}\,  \Vert V_\lambda\Vert_{C([0,\tau] \times \overline{\Omega})} =\infty$, where $V_\lambda$ satisfies \eqref{3.42}.
 As in \cite[Theorem 1.1]{DLG2018}, if we assumed $\lim_{\lambda \uparrow\Sigma(\infty,c)}\,  \Vert V_\lambda\Vert_{C([0,\tau] \times \overline{\Omega})} <\infty$, there would exist an increasing sequence $\{\lambda_n\}^\infty_{n=1}$ with $\lambda_n>\Sigma(0,c)$ and $M>0$ such that
$$ \lim_{n\to \infty}\,\lambda_n=\Sigma(\infty,c)< \infty\ \textrm{and}\ V_{\lambda_n}< V_{\lambda_{n+1}}\leq M\, \textrm{in}\, [0,\tau]\times \overline{\Omega}\, \textrm{ for all}\, \ n\geq 1, $$
whence follows that the pointwise limit
\begin{eqnarray} \label{sd}
V_\infty(t,x):=\lim_{n\to \infty} V_{\lambda_n}(t,x)\, \textrm{in}\, [0,\tau]\times \overline{\Omega}\,
\end{eqnarray}
exists and is bounded from below by $V_{\lambda_1}$.

Since
\begin{eqnarray} \label{se} 0<V_{\lambda_n}=(\frac{\partial}{\partial t}-d\Delta )^{-1}[(\lambda_n+ a(t,x)) V_{\lambda_n}-b(t,x)V^p_{\lambda_n}],\ \mbox{for every } n\geq 1,
\end{eqnarray}
 is  uniformly bounded from above by $M$, and $(\frac{\partial}{\partial t}-d\Delta )^{-1}$ can be regarded as a compact operator in the space
\begin{eqnarray*}
\begin{array}{ll}
{D}_0([0, \tau]\times \overline \Omega) =&:=\{\xi(t,x)\in {C}([0, \tau]\times \overline \Omega): \ \xi(t,x)=0\ \forall (t,x)\in [0, \tau]\times \partial \Omega, \\
&\xi(0,x)=c \xi(\tau,x),\ \forall x\in \overline \Omega\},
\end{array}
\end{eqnarray*}
endowed with the norm $\Vert \cdot \Vert_{C([0,\tau] \times \overline{\Omega})}$, we can infer that the pointwise limit in \eqref{sd} is a  uniform limit.

So, by letting $n\to \infty$ in \eqref{se} yields that $V_\infty$ is a positive solution for problem \eqref{3.42} with $\lambda=\Sigma(\infty,c)$. Equivalently,
$U_\infty$, given by \eqref{e38} and  \eqref{e39} is a positive solution of \eqref{p351} with $\lambda=\Sigma(\infty,c)$, which leads a contradiction to the item $(iii)$ of Corollary \ref{ap2}. This ends the proof.
\epf

\section{The dynamical behavior of the solution}

Now, we are in position to prove the dynamical behavior of the unique global classical solution $u(t,x)\in C^{1+\theta/2, 2+\theta}((n\tau, (n+1)\tau]\times \Omega)\bigcup C((n\tau, (n+1)\tau])\times\overline \Omega)$ $(n=0,1,2,\cdots)$. As we already know, it was established in \cite{MGL} for any nonnegative and continuous initial value $u_0$ that vanishes on the boundary $\partial \Omega$.

\begin{thm}\label{t41} For any nonnegative and nontrivial function $u_0\in C(\overline \Omega)$ given, the solution $u(t,x)$ of problem \eqref{a01}  behaves in the following way:
\begin{enumerate}
    \item[$i)$] If either $b(t,x)\not\equiv 0$ in $[0,\tau]\times \overline\Omega$  and $\Sigma (0,c)\geq 0$ or $b(t,x)\equiv 0$  in $[0,\tau]\times \overline\Omega$ and $\Sigma (0,c)> 0$, then $\lim_{t\to \infty} u(t,x)=0$ uniformly for $x\in \overline\Omega$;
     \item[$ii)$] if $\Sigma(0,c)<0<\Sigma(\infty,c)$, then $\lim_{m\to \infty} u(t+m\tau,x)=U(t,x)$ for any $t\geq 0$ and uniformly for $x\in \overline \Omega$, where $U(t,x)$ is the unique solution of the problem \eqref{a012};
      \item[$iii)$] If either $b(t,x)\not\equiv 0$ in $[0,\tau]\times \overline\Omega$  and $\Sigma (\infty,c)\geq 0$ or $b(t,x)\equiv 0$  in $[0,\tau]\times \overline\Omega$ and $\Sigma (\infty,c)<0$, then $\lim_{t\to \infty} ||u(t,\cdot)||_{C(\overline \Omega)}=+\infty$.
\end{enumerate}
\label{abj}
\end{thm}
\bpf
 We first proved the part $(ii)$. Without loss of generality, let us assume that $u_0(x)>0$ for $x \in \Omega$; otherwise, we may use the initial time $t_0>0$ replaced with $t=0$. Since we are assuming $\Sigma(0,c)<0<\Sigma(\infty,c)< \infty$, we can choose the upper solution $\overline U$, defined in \eqref{d32}, and the lower solution $\underline U$, defined in \eqref{4.52}. Owing to their definitions and the fact that the eigenvalue functions $\phi$ and $\varphi$ are strongly positive, a sufficiently small $\varepsilon >0$ and a large $K>1$ can be selected to lead to the below inequalities
$$\underline U (0,x) = \varepsilon\phi(0,x)\leq{u}(0,x)\leq K \varphi(0,x) = \overline U (0,x),\ \mbox{for } x\in \overline \Omega,$$ and
 $$\underline U(0^+,x)={c \underline U (0,x)}\leq c{u}(0,x)={u}(0^+,x)\leq K \varphi (0^+,x) = \overline U(0^+,x)\quad \mbox{for }x\in \overline \Omega.$$

 So, it follows from the comparison principle that $\underline U(t,x)\leq{u}(t,x)\leq\overline U(t,x)$ for $t\in(0^+,\tau]$ and $x\in \overline \Omega$. By using a similar process for $t\in (\tau^+,2\tau], ((2\tau)^+,3\tau],\cdots$, we infer that
\begin{equation}\label{i40}
    \underline U (t,x)\leq{u}(t,x)\leq \overline U(t,x),\,\,t\geq 0,\,\,x\in \overline \Omega.
\end{equation}

Below, let us  construct two iteration sequences, say $\{\bar u^{(n)}\}$ and $\{\underline u^{(n)}\}$, starting from $\underline u^{(0)}:= \underline U$ and $\bar u^{(0)}:= \overline U$, respectively, defined by
\begin{eqnarray}
\left\{
\begin{array}{lll}
\underline{u}_t^{(n)}-d\Delta\underline{u}^{(n)}+b^M\underline{u}^{(n)}=b^M\underline{u}^{(n-1)}+a(t,x) \underline{u}^{(n-1)}&\\
-b(t,x)(\underline{u}^{(n-1)})^p,& t\in(0^+,\tau],\,\,x \in \Omega,  \\[1mm]
\underline u^{(n)}(t,x)=0, & t\in[0,\tau],\,\,x\in\partial\Omega\\
\underline u^{(n)}(0,x)=\underline u^{(n-1)}(\tau,x),&  x\in \overline\Omega,\\
\underline u^{(n)}(0^+,x)=c \underline u^{(n-1)}(0,x),& \ x\in\Omega,
\end{array} \right.
\label{p42}
\end{eqnarray}
and
\begin{eqnarray}
\left\{
\begin{array}{lll}
\bar{u}_t^{(n)}-d\Delta\bar{u}^{(n)}+b^M\bar{u}^{(n)}=b^M \bar{u}^{(n-1)}+a(t,x) \bar{u}^{(n-1)}&\\
-b(t,x)(\bar{u}^{(n-1)})^p,& t\in(0^+,\tau],\,\,x\in \Omega,  \\[1mm]
\underline u^{(n)}(t,x)=0, & t\in[0,\tau],\ \ x\in\partial\Omega\\
\bar u^{(n)}(0,x)=\bar u^{(n-1)}(\tau,x), & \ x\in \overline\Omega,\\
\bar u^{(n)}(0^+,x)=c \bar u^{(n-1)}(0,x),& x\in\Omega,
\end{array} \right.
\label{p43}
\end{eqnarray}
where $b^M>0$ is taken as $b^M=\max_{[0,\tau]\times \overline \Omega}[-a(t,x)+p b(t,x) (\overline U)^{p-1}]$ to ensure the monotonicity of the function $h(x,s):=b^M s+a(t,x) s-b(t,x)s^p$ with respect to $s>0$.

From \eqref{i40}, \eqref{p42} and \eqref{p43}, we have
$$\underline{u}^{(1)}(0,x)=\underline{u}^{(0)}(\tau,x)\leq{u(\tau,x)}
\leq\bar{u}^{(0)}(\tau,x)=\bar{u}^{(1)}(0,x),\\
$$
for $x\in \overline \Omega$,
and
$$\underline{u}^{(1)}(0^+,x)=c\underline{u}^{(0)}(\tau,x)\leq{c u(\tau,x)}
=u(\tau^+,x)\leq{c \bar{u}^{(0)}(\tau,x)}=\bar{u}^{(1)}(0^+,x),\\
$$
for $x\in \overline \Omega$.
Thus, by comparison argument, we have
$$\underline{u}^{(1)}(t,x)\leq{u(t+\tau,x)}\leq{\bar{u}^{(1)}(t,x)},\ t\in(0^+,\tau]\ \mbox{and }x\in \overline \Omega,$$
hold so that a induction process leads to
$$\underline{u}^{(1)}(t,x)\leq{u(t+\tau,x)}\leq{\bar{u}^{(1)}(t,x)} \ \mbox{for } t\geq 0\ \mbox{and } x\in\overline \Omega.$$
Similarly, we can conclude by iteration that
\begin{equation}\label{i44}
\underline{u}^{(m)}(t,x)\leq{u(t+m\tau,x)}\leq{\bar{u}^{(m)}(t,x)},\,t\geq{0},\,x\in\overline \Omega
\end{equation}
hold for any $m \in \mathbb{N}$.

Besides this, it is standard to show that
\begin{equation}\label{i45}
\underline{u}^{(m)}(t,x)\leq{\underline{u}^{(m+1)}(t,x)},\ \mbox{and }\overline{u}^{(m+1)}(t,x)\leq{\overline{u}^{m}(t,x)}\ \mbox{for all }t\geq{0},\,x\in\overline \Omega,
\end{equation}
hold for any $m \in \mathbb{N}$. So, it follows from \eqref{i44} and \eqref{i45} that the limits
$$\lim\limits_{m\to\infty}{\underline{u}^{(m)}(t,x)}\ \mbox{and }
\lim\limits_{m\to\infty}{\bar{u}^{(m)}(t,x)},\ t\in(0^+,\tau]\ \mbox{and }x\in \overline \Omega$$
exist, and again is standard showing that they are the minimal and the maximal $\tau$-periodic solutions of problem \eqref{a012}, respectively, so that they are equals due to the uniqueness of solutions of problem \eqref{a012} stated in Theorem \ref{ap}-$(iii)$. This completes the proof of $(ii)$.

Below, let us begin to prove the part $(iii)$. To do this, first we point out that the arguments just used to prove the item $(ii)$ above, can be repeated to prove the following Lemma.
\begin{lem}\label{l42}
If $\Sigma(0,c)<\lambda<\Sigma(\infty,c)$, then $\lim_{m\to \infty} u_{\lambda}(t+m\tau,x)=U_{\lambda}(t,x)$ for any $t\geq 0$ and uniformly for $x\in \overline \Omega$, where $U_{\lambda}$ is the unique solution of the problem \eqref{p351}, and $u_{\lambda}$ is a unique positive solution of
\begin{eqnarray}
\left\{
\begin{aligned}
&u_t=d \Delta u+ a(t,x) u-b(t,x)u^p + \lambda u, && t\in((n\tau)^+,(n+1)\tau],\,\,x \in \Omega,  \\
&u(t,x)=0, && t>0,\,\,x \in \partial \Omega,\\
&u(0,x)=u_0(x), && x\in \overline \Omega,\\
&u((n\tau)^+,x)=cu(n\tau,x),&& x\in \overline \Omega,\ n=0,1,2,\cdots.
\end{aligned} \right.
\label{a011}
\end{eqnarray}
\end{lem}

Coming back to the prove of item $(iii)$, we first note that the solution $u$ of problem \eqref{a01} satisfies, by comparison principle, the inequality
$$u_\lambda(t,x)\leq  u(t,x)\ \mbox{in }  [0,\infty)\times \overline \Omega,$$
 for all $\Sigma(0,c)< \lambda< \Sigma(\infty,c) \leq 0$ given, where $u_\lambda$ is the unique solution of problem \eqref{a011}.

So, by Lemma \ref{l42}, we have that
$$U_\lambda(t,x)=\lim_{m\to \infty}\, u_\lambda (t+m\tau,x)\leq \liminf_{t\to \infty}\, u (t+m\tau,x),$$
for any $\Sigma(0,c)< \lambda< \Sigma(\infty,c)$, $t\geq 0$ and uniformly for $x\in \overline \Omega$.

So, by letting $\lambda \uparrow \Sigma(\infty,c)$, together with \eqref{ma2}, yields
$$\liminf_{t\to \infty} ||u(t,\cdot)||_{C(\overline \Omega)}=\infty,$$
which concludes the proof of $(iii)$.

The proof of the item $(i)$ is similar to that one used to prove the item $(iii)$,  by using \eqref{ma1}, and we omit the detail with obvious modification.
\epf

As a consequence of Theorems \ref{abj} and \ref{t25}, we have the next result that highlights the impact of the impulse on the dynamics of the population.

\begin{thm} Assume that $b(t,x)\not\equiv 0$ in $[0,\tau]\times\Omega$ and $d$, $a(t,x)$ and $b(t,x)$ are fixed. Then there exist $0<c_*<\infty$ and $c_*< c^* \leq \infty$ such that the unique solution $u(t,x)$ of problem \eqref{a01} behaves in the following way:
\begin{enumerate}
    \item[$i)$] if $0<c \leq c_*$, then $\lim_{t\to \infty} u(t,x)=0$ uniformly for $x\in \overline\Omega$;
    \item[$ii)$] if $c_*<c<c^*$, then $\lim_{m\to \infty} u(t+m\tau,x)=U(t,x)$ uniformly for $x\in  \overline \Omega$, where $U(t,x)$ is the unique solution of the problem \eqref{a012};
    \item[$iii)$] if $c \geq c^*$,  then $\lim_{t\to \infty} ||u(t,\cdot)||_{C(\overline \Omega)}=+\infty$.
\end{enumerate}
\label{abm}
\end{thm}
\bpf It follows from Theorem \ref{t24}
that $\Sigma (0, z)$ is continuous and strictly decreasing with respect to $z>0$. Moreover, we know that $\lim_{z\to\infty} \Sigma(0,z)=-\infty$ and $\lim_{z\to 0^+} \Sigma(0,z)=+\infty$. Therefore, there exists uniquely $c_*>0$ such that $\Sigma (0, c_*)=0$, and $\Sigma (0, c)\geq 0$ if, and only if, $c \leq c_*$, which means, by Theorem \ref{abj}-$(i)$, that  $\lim_{t\to \infty} u(t,x)=0$ uniformly for $x\in \overline\Omega$ if $c \leq c_*$, that completes the proof of $(i)$.

 Next, let us set $c^*>0$ to prove the items $(ii)$ and  $(iii)$, that is, let us set a $c_*<c^* \leq \infty$ such that $\Sigma (\infty, c) >0$ for any $c<c^*$. By the assumption $b(t,x)\not\equiv 0$ in $[0,\tau]\times\Omega$, we have from Theorem \ref{t23}-$(iv)$, that $\Sigma (0, z) < \Sigma (\infty, z) \leq \infty$ for each $z>0$. Besides this, it is a consequence of Theorem \ref{t25} that just one situation may occur: either $(a)$ $\Sigma (\infty, z) = \infty$ for any $z>0$ or  $(b)$ $\Sigma (\infty, z) < \infty$ for any $z>0$.

 If the possibility $(a)$ occurs, then $\Sigma (\infty, z) = \infty$ for any $z>0$ that naturally leads us to set $c^* = \infty$, that is, $\Sigma(\infty,c)=\infty$ for all $c>0$. This case is similar to the classical logistic system, that is, unlimited growth does not occur.

Assume that the possibility $(b)$ holds true. Then Theorem \ref{t25} implies that there exists a $c^*>0$ such that $\Sigma (\infty,c^*)=0$, and $\Sigma (\infty, c)\leq 0$ if, and only if, $c \geq c^*$. Since $\Sigma(0, c^*) < \Sigma(\infty,c^*)=0$, we have $c_*<c^*$.  So, for all $c \geq c^*$, we have  by Theorem \ref{t41}-$(iii)$ that the item $(iii)$ above occurs so that unlimited growth happens.

To finish the prove, we just note that the above construction of  $0<c_* < c^* \leq \infty$ implies that  $\Sigma(\infty,c)<0<\Sigma(\infty, c)$ holds if, and only if, $c_* < c<c^*$ so that the claim of the item $(ii)$ above follows from Theorem \ref{t41}-$(ii)$. This ends the proof.
\epf

The above analysis can be immediately replicated to the case $b(t,x)\equiv 0$ in $[0, \tau]\times \overline \Omega$ by just noting that  $\Sigma (0, z)=\Sigma (\infty, z)$ for any $z>0$ whence follows  that $c_*=c^*$. So, we have the next result that extends some classical results for Malthusian model.

\begin{thm}  Assume that $b(t,x)\equiv 0$ in $[0, \tau]\times \overline \Omega$, and $d$ and $a(t,x)$ are fixed. Then there exists $0<c_*< \infty$ such that the unique solution $u(t,x)$ of problem  \eqref{a01} satisfies:
\begin{enumerate}
    \item[$i)$]  $\lim_{t\to \infty} u(t,x)=0$ uniformly for $x\in \overline\Omega$ if $c<c_*$;
    \item[$ii)$] $\lim_{t\to \infty} ||u(t,\cdot)||_{C(\overline \Omega)}=+\infty$ if $c >c_*$;
     \item[$iii)$] $k\phi(t,x)$ is the positive solution to problem  \eqref{a01} for any $k>0$ if $c =c_*$, where $(c_*,\phi(t,x)$ is the eigen-pair of the eigenvalue problem \eqref{a013}. 
\end{enumerate}
\label{abm1}
\end{thm}
\bpf Similarly as in Theorem \ref{abm}, there exists uniquely $c_*>0$ such that $\Sigma (0, c_*)=0$ since $\Sigma (0, z)$ is continuous and strictly decreasing with respect to $z>0$, and $\lim_{z\to\infty} \Sigma(0,z)=-\infty$ and $\lim_{z\to 0^+} \Sigma(0,z)=+\infty$. Therefore, if
 $c<c_*$, then $\lim_{t\to \infty} u(t,x)=0$ uniformly for $x\in \overline\Omega$ by Theorem \ref{abj} $(i)$.

 On the other hand, the assumption that $b(t,x)\equiv 0$ in $[0, \tau]\times \overline \Omega$ assures that $\Sigma (0, c)=\Sigma (\infty, c)$ for any $c>0$. Therefore,  $\Sigma (0, c_*)=\Sigma (\infty, c_*)=0$, which means that, if $c>c_*$, then $\Sigma (\infty, c)<0$ and $\lim_{t\to \infty} ||u(t,\cdot)||_{C(\overline \Omega)}=+\infty$, unlimited growing happens.
 
 The result of (iii) is from the definition of the eigenvalue value in the case that $b(t,x)\equiv 0$ in $[0, \tau]\times \overline \Omega$. 
\epf

\section{Numerical simulation and biological explanation}

In this section, we consider the interval $\Omega=[0, \pi]$, and we will assume that the impulse occurs at each time $\tau =2$. When $0<c<1$, impulsive harvesting will take place, while $c>1$, birth pulse may occur.

We now fix some parameters in the  model \eqref{a01}. Set $d=1$, $a(t,x)=2.0$, the initial value  $u_0(x)=0.5\sin{x}+0.2\sin{3x}$, and $b(t,x):=b(x)=[0.01+0.5\sin(3x)]^+$ for $x\in [0, \pi]$ and any $t\geq 0$, which means that $b(x)\equiv 0$ for $x\in [\frac \pi 3-\frac13\arcsin{0.02}, \frac {2\pi} 3-\frac13\arcsin{0.02}]$.

\begin{figure}[H]
\centering
\subfigure[]{ {
\includegraphics[width=0.35\textwidth,height=3.5cm]{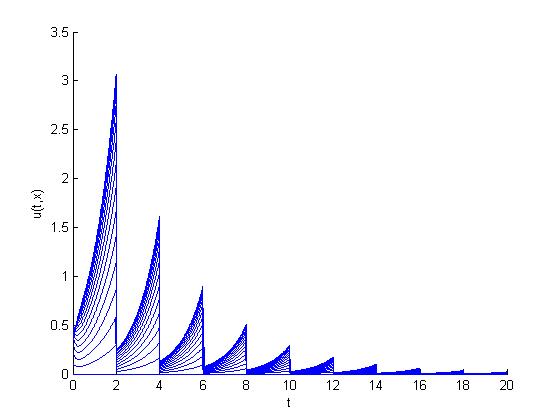}
} }
\subfigure[]{ {
\includegraphics[width=0.35\textwidth,height=3.5cm]{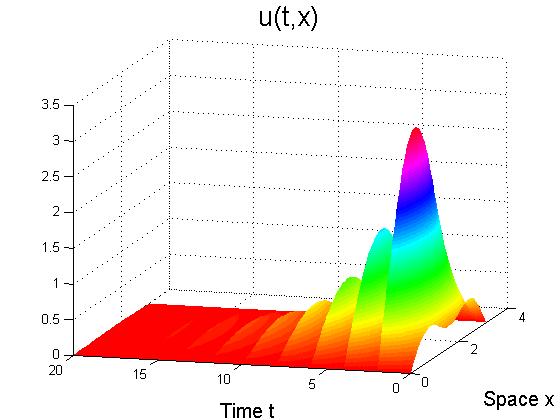}
} }

\subfigure[]{ {
\includegraphics[width=0.35\textwidth,height=3.5cm]{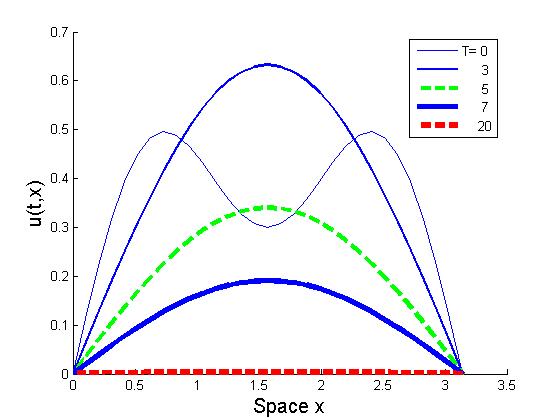}
} }
\caption{\scriptsize The dynamics of species $u$ with small pulse $c=0.08$ at every $\tau=2$.  Graph $(a)$ is the right view of the spatiotemporal distribution of $u$ plotted in graph $(b)$, while in the Graph $(c)$ are plotted the cross sections view at $t=0,3,5,7, 20$.  Graphs $(a)-(c)$ imply that the species $u(t,x)$ gradually tends to zero with time $t$.}
\label{tu1}
\end{figure}

First, by choosing  small impulse $c=0.08$, we see in the Fig. \ref{tu1} that the species goes to the extinction quickly. This was proved in Theorem \ref{abm}-$i)$.

%figure1

\begin{figure}[H]
\centering
\subfigure[]{ {
\includegraphics[width=0.45\textwidth,height=4cm]{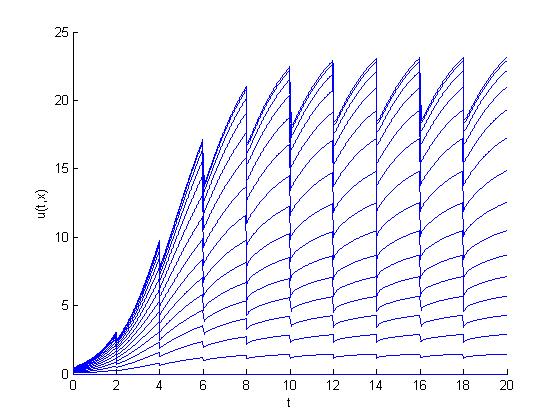}
} }
\subfigure[]{ {
\includegraphics[width=0.45\textwidth,height=4cm]{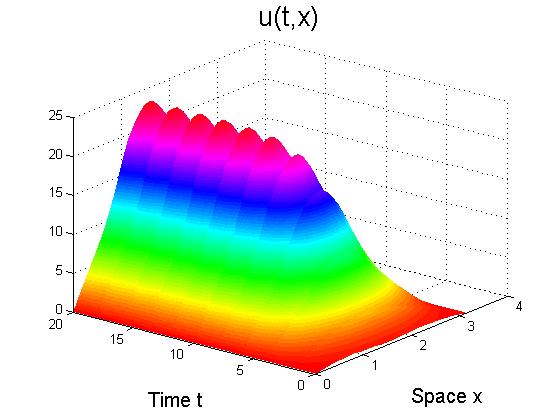}
} }

\subfigure[]{ {
\includegraphics[width=0.45\textwidth,height=4cm]{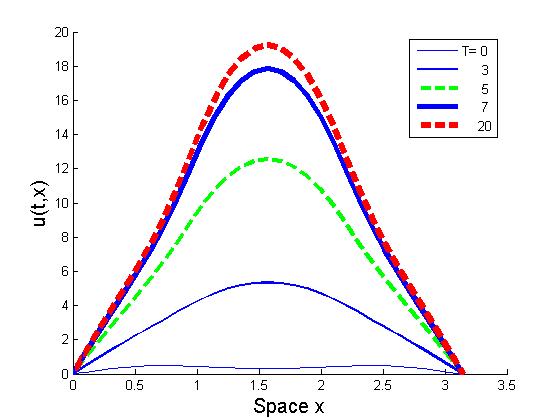}
} }
\caption{\scriptsize The dynamics of species $u$ with medium pulse $c=0.8$ at every $\tau=2$.  Graph $(a)$ is the right view of the spatiotemporal distribution of $u$ in the graph $(b)$, and Graph $(c)$ is its cross section views at $t=0,3,5,7, 20$.  Graphs $(a)- (c)$ imply that the species $u(t,x)$ gradually tends to a periodic state.}
\label{tu2}
\end{figure}

Secondly, let us choose $c=0.8$. Figure \ref{tu2} shows that the solution stabilizes to a periodic positive
state. See Theorem \ref{abm}-$ii)$.

%figure2

\begin{figure}[H]
\centering
\subfigure[]{ {
\includegraphics[width=0.45\textwidth,height=4cm]{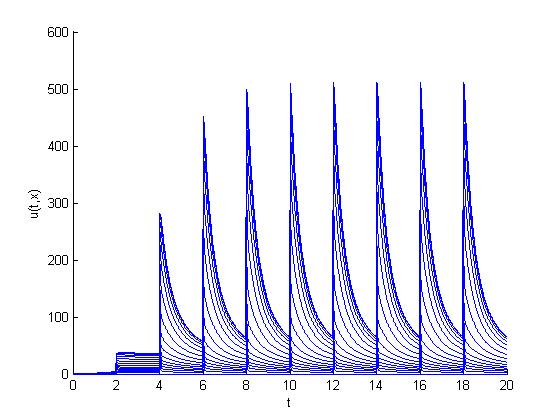}
} }
\subfigure[]{ {
\includegraphics[width=0.45\textwidth,height=4cm]{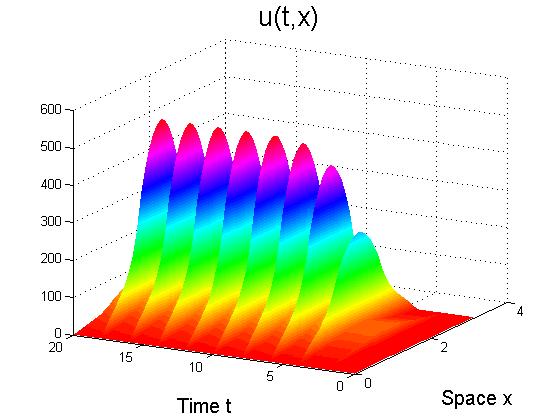}
} }

\subfigure[] {
\includegraphics[width=0.45\textwidth,height=4cm]{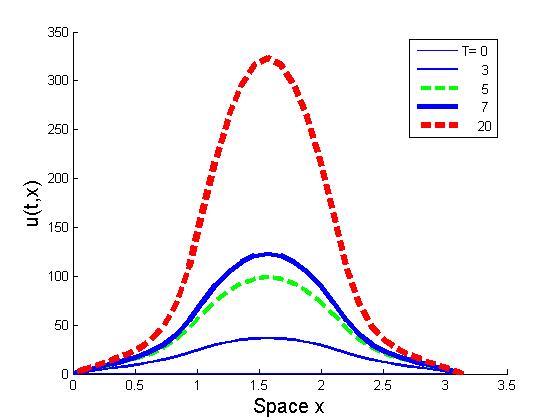}}

\caption{\scriptsize The dynamics of species $u$ with large pulse $c=0.08$ at every $\tau=2$.  Graph $(a)$ is the right view of the spatiotemporal distribution of $u$ in graph $(b)$, and Graph $(c)$ is its cross section views at $t=0,3,5,7, 20$.  Graphs $(a)- (c)$ imply that the species $u(t,x)$ increases quickly.
}
\label{tu3}
\end{figure}

Lastly, we choose a large impulsive rate $c=7.8$, it follows from Theorems \ref{abm}-$iii)$ that the solution grows quickly, see Figure \ref{tu3}.
%figure3

To sum up, the rate of population extinction will be accelerate with  small impulses,  medium-impulse rates are more favorable for species to persist; however, large
impulses make the population increases quickly and eventually blows up.

It is worth mentioning that all above results hold if $-\Delta$ operator in model \eqref{a01} is replaced with a second-order uniformly strongly elliptic operator with obvious modifications.

\end{document}